\documentstyle[aas2pp4,epsf]{article}

\def\simless{\mathbin{\lower 2pt\hbox
   {$\rlap{\raise 5pt\hbox{$\char'074$}}\mathchar"7218$}}}   
\def\simgreat{\mathbin{\lower 2pt\hbox
   {$\rlap{\raise 5pt\hbox{$\char'076$}}\mathchar"7218$}}}   

\begin{document}

\title{Collapse of a Molecular Cloud Core to Stellar Densities:\\
The First Three-Dimensional Calculations}

\author{Matthew R. Bate}

\affil{Max-Planck-Institut f\"ur Astronomie, K\"onigstuhl 17, D-69117 Heidelberg, Germany}

\begin{abstract}

We present results from the first three-dimensional calculations 
ever to follow the collapse of a molecular cloud core 
($\sim 10^{-18} {\rm \ g \ cm}^{-3}$) to stellar densities
($> 0.01 {\rm \ g \ cm}^{-3}$).
The calculations resolve structures over 7 orders of magnitude in 
spatial extent ($\sim 5000\ {\rm AU} - 0.1\ {\rm R}_\odot$), and over 
17 orders of magnitude in density contrast.  
With these calculations, 
we consider whether fragmentation to form a close binary stellar system
can occur during the second collapse phase.  We find that, if the 
quasistatic core that forms before the second collapse phase is dynamically
unstable to the growth of non-axisymmetric perturbations, the 
angular momentum extracted from the central regions of the 
core, via gravitational torques, is sufficient to prevent 
fragmentation and the formation of a close binary during 
the subsequent second collapse.

\end{abstract}


\section{Introduction}

The formation of close binary stellar systems is an, as yet, unsolved 
problem in the field of star formation.  Calculations following the isothermal
collapse and fragmentation of molecular cloud cores can successfully
produce binaries with separations $\simgreat 1$ AU (e.g.\ Boss 1986).
However, once a collapsing molecular cloud core becomes optically 
thick, the gas temperature increases with density, a hydrostatic core is 
formed with radius $\sim 4$ AU (\cite{Larson69}), and fragmentation is
suppressed (Boss 1986, 1988).

A potential opportunity to form close binaries via fragmentation occurs
when the temperature in the hydrostatic core reaches $\sim 2000$ K and 
molecular hydrogen begins to dissociate, resulting in a second collapse
within the hydrostatic core (\cite{Larson69}).  When the dissociation is
complete, the collapse is finally stopped with the formation of a 
second hydrostatic, or stellar, core.  Fragmentation during the
second collapse has been obtained by Boss (1989) and Bonnell \& Bate (1994), 
although in the former paper the binary quickly merged.  In both papers, 
fragmentation was found to occur only if the ratio of the rotational
energy to the magnitude the gravitational potential energy, $\beta$, 
exceeded the value required for dynamic instability to non-axisymmetric 
perturbations ($\beta \simgreat 0.274$; Durisen et al.\ 1986) 
after a stellar core formed.  Otherwise, only a stellar core and stable
disc form.  However, in both papers, only the inner regions of the first 
hydrostatic core were modelled and the initial conditions were chosen 
somewhat arbitrarily.
Ideally, to study the formation of close binary systems, one would 
wish to follow the collapse of an optically-thin, molecular cloud core all the
way to stellar densities.  However, such a calculation requires 
resolving densities from $10^{-18} - 0.01{\rm\ g \ cm}^{-3}$ and
length scales from $\sim 10^{17} - 10^{10}$ cm in three dimensions.

To achieve such high spatial resolution in three dimensions we use the 
Lagrangian smoothed particle hydrodynamics (SPH) method.  With 
SPH, the spatial resolution is given by the smoothing length 
(over which hydrodynamical properties are calculated) which in 
modern implementations is continuously variable (e.g.\ Benz et al.\ 1990).  
The smoothing lengths are set by ensuring that each particle contains a 
certain number of neighbours, $N_{\rm neigh}$, or equivalently a fixed
mass, within two smoothing lengths.  Thus, in contrast to a
grid-based code which has spatially-limited resolution, SPH has mass-limited 
resolution which {\it automatically} gives greater spatial resolution in 
regions of higher density.
Bate \& Burkert (1997) recently demonstrated this mass-limited resolution. 
They found that to follow the collapse and fragmentation of a Jeans-unstable 
gas cloud reliably, the minimum resolvable mass should always be 
less than the Jeans mass (see also Whitworth 1998).  
In practice, this means that a Jeans mass should always 
be represented by at least $\approx 2 N_{\rm neigh}$ particles.  

The mass-limited resolution of SPH is ideal for studying the collapse 
and fragmentation of molecular cloud cores because there is a minimum 
Jeans mass in the problem (Figure 1).  By contrast,
there is no minimum Jeans length.  This is a problem for grid-based codes
which must resort to nested or adaptive grids (e.g.\ Burkert \& 
Bodenheimer\ 1993; Truelove et al.\ 1998).  With SPH, if the number of 
particles used is sufficient to resolve the minimum Jeans mass, a 
calculation can be followed to arbitrary densities with the required
spatial resolution given automatically with increasing density.  

\begin{figure}[t]
\plotone{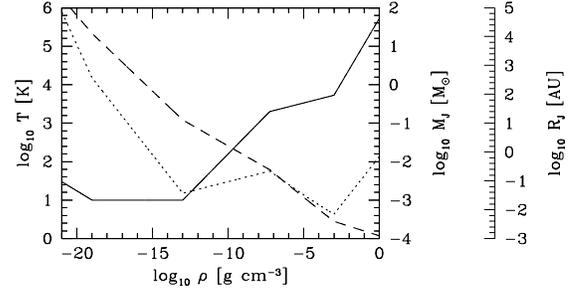}
\caption{The gas temperature $T$ (solid line), Jeans mass $M_{\rm J}$ (dotted line), and Jeans radius $R_{\rm J}$ (dashed line), as a function of density in a collapsing molecular cloud core (adapted from Tohline 1982).  The minimum Jeans mass of $ \approx 4 \times 10^{-4}~{\rm M}_{\odot}$ occurs at the end of the second collapse phase when the dissociation of molecular hydrogen is almost completed at a density of $\sim 10^{-3} {\rm \ g \ cm}^{-3}$.}
\end{figure}

\section{Method}

The calculations were performed using a three-dimensional SPH
code based on a version originally 
developed by Benz (\cite{Benz90,Benzetal1990}).  The code uses a tree
to calculate the gravitational forces and find nearest neighbours.
The smoothing lengths are variable in time and space, subject to 
the constraint that the number of neighbours of each particle must 
remain approximately constant
at $N_{\rm neigh} = 50$.  The gravitational force between particles
is softened over the hydrodynamic smoothing length using spline 
softening derived from the smoothing kernel.
Hence, the gravitational softening length and hydrodynamic
smoothing length are equal and decrease with increasing density.
The standard form of the artificial viscosity is used 
(\cite{MonGin83}), with the parameters 
$\alpha_{\rm v}=0.5$ and $\beta_{\rm v}=1.0$.  
The SPH equations are integrated using a second-order 
Runge--Kutta--Fehlberg integrator. 
Individual time-steps are used for each particle (\cite{BBP95}).  
Further details can be found in the above references.

The code does not include radiative transfer.  Instead, 
to model the behaviour of the gas during the different phases of collapse,
we use a piece-wise polytropic equation of state $P=K \rho^{\gamma}$, 
where $P$ is 
the pressure, $\rho$ is the density, $K$ gives the entropy of the gas, and
the ratio of specific heats, $\gamma$, is varied as
\begin{equation}
\label{polytropic}
\gamma = \cases {\begin{array}{lll}
1	& 			  & \rho \leq 1.0 \times 10^{-13} \cr
7/5 	& \ 1.0 \times 10^{-13}  < \hspace{-6pt} & \rho \leq 5.7 \times 10^{-8} \cr
1.15 	& \ 5.7 \times 10^{-8\ } < \hspace{-6pt} & \rho \leq 1.0 \times 10^{-3} \cr
5/3 	& 			  & \rho > 1.0 \times 10^{-3} \cr
\end{array}}
\end{equation}
where the densities are in ${\rm g\ cm}^{-3}$ (see Figure 1).  
The values of $\gamma$ and the transition densities 
are derived from Tohline (1982).  The variable value of $\gamma$ 
mimics the following behaviour of the gas.  The
collapse is isothermal ($\gamma=1$) until the gas becomes optically thick 
to infrared radiation 
at $\rho \approx 10^{-13} {\rm \ g \ cm}^{-3}$, beyond which $\gamma=7/5$
(appropriate for a diatomic gas).
When the gas reaches a temperature of $\approx$ 2000 K 
($\rho = 5.7 \times 10^{-8} {\rm \ g \ cm}^{-3}$), molecular 
hydrogen begins to dissociate and the temperature only slowly increases
with density.  In this phase we use $\gamma = 1.15$ to 
model {\it both} the decreasing mean molecular weight and the slow increase 
of temperature with density, the latter of which has an effective 
$\gamma \approx 1.10$.
Finally, when the gas has fully dissociated
($\rho \approx 10^{-3} {\rm \ g \ cm}^{-3}$), the gas is monatomic and
$\gamma=5/3$.  The value of $K$ is
defined such that when the gas is isothermal, $K = c_{\rm s}^2$
with $c_{\rm s} = 2.0 \times 10^4 {\rm \ cm\ s^{-1}}$, and when $\gamma$
changes the pressure is continuous.

To test that the above equation of state captures the important elements
of the gas's behaviour, we have performed spherically-symmetric,
one-dimensional (1-D), finite-difference calculations of the collapse of a
molecular cloud core using the above equation of state
and obtained excellent agreement with results from 1-D
calculations incorporating radiative transfer 
(e.g.\ Larson 1969; Winkler \& Newman 1980a, b).  We have also tested that the 
three-dimensional (3-D) SPH code can indeed accurately resolve the 
collapse down to stellar densities by performing comparison
calculations with the same initial conditions as used by the 1-D 
finite-difference calculations.  So long as the number of particles used by an
SPH calculation is sufficient to resolve the Jeans mass throughout the
calculation (see below),
we obtain excellent agreement with the results from the 1-D 
finite-difference code.  The details of these test calculations
will be presented in a later paper.

\section{Calculations}

We present results from a 3-D
SPH calculation following the collapse 
of an initially uniform-density, uniform-rotating, molecular cloud 
core of mass $M=1 {\rm \ M}_\odot$ and
radius $R=7 \times 10^{16} {\rm \ cm}$.  The angular frequency is
$\Omega=7.6 \times 10^{-14}$ rad s$^{-1}$, and the ratios of the thermal
and rotational energies to the magnitude of the gravitational 
potential energy are $\alpha = 0.54$ and $\beta=0.005$, respectively.
To satisfy the resolution criterion of Bate \& Burkert (1997),
we require that the minimum Jeans mass during the calculation
($\approx 4 \times 10^{-4}\ {\rm M}_{\odot}$) contains at least 
$\approx 2 N_{\rm neigh} = 100$ particles.  
Hence, we use $3 \times 10^5$ equal-mass particles.

\begin{figure}[t]
\epsscale{1.00}
\plotone{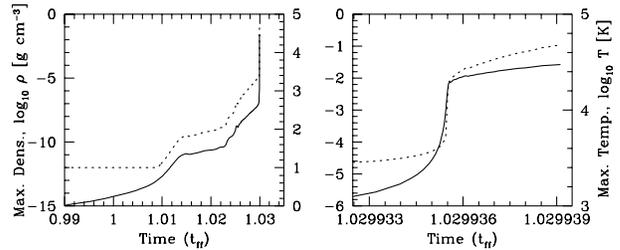}
\caption{Maximum density (solid line) and maximum temperature (dotted line) versus time for the collapsing molecular cloud core.  Time is given in units of the initial free-fall time ($t_{\rm ff}=1.785 \times 10^{12}$ s).  The right graph has expanded axes to show the second collapse phase in greater detail.}
\end{figure}

\begin{figure*}[t]
\epsscale{2.00}
\plotone{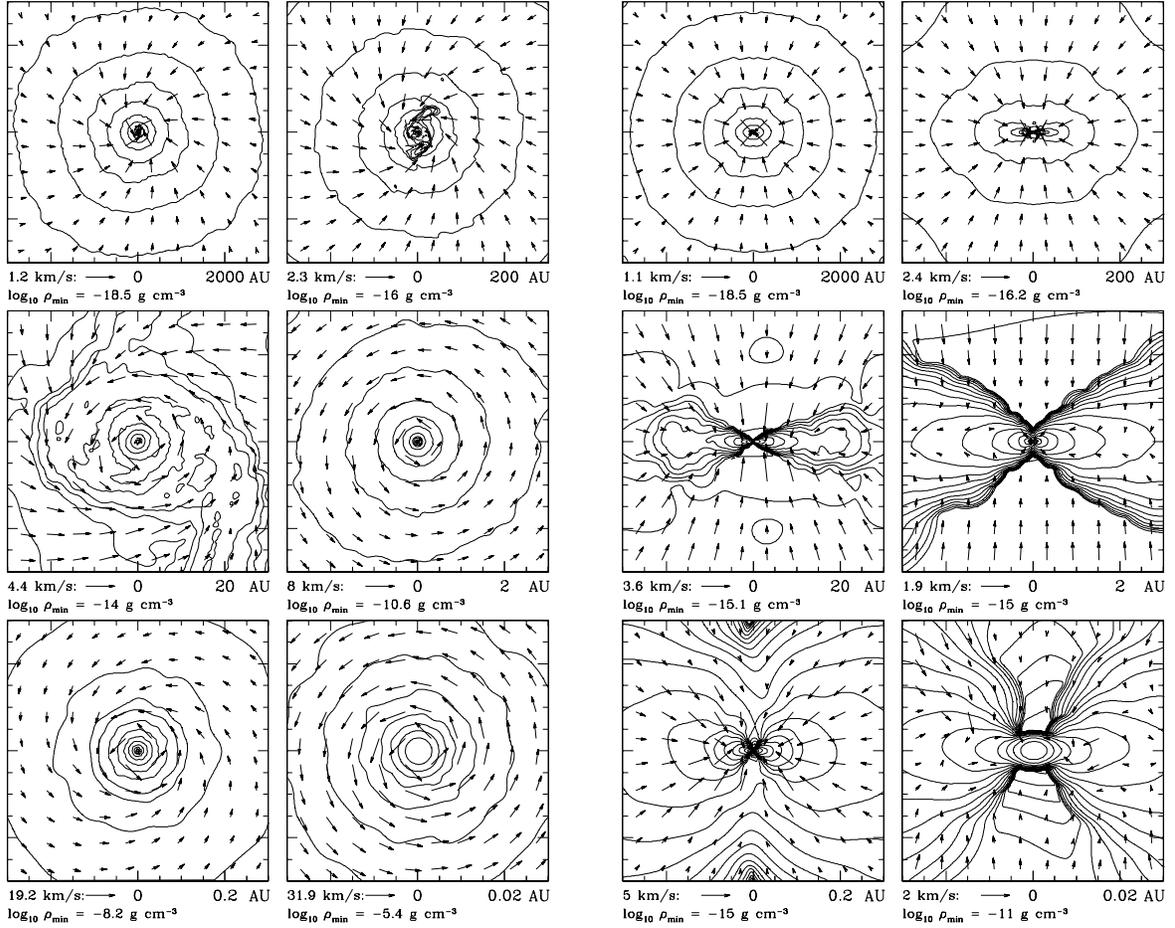}
\caption{The state of the system at the end of the calculation.  The six panels on the left give the density and velocity in the plane perpendicular to the rotation axis and through the stellar core.  The six panels on the right give the density and velocity in a section down the rotation axis.  In each case, the six consecutive panels give the structure on a spatial scale that is 10 times smaller than the previous panel to resolve structure from 3000 AU to $\approx 0.2~{\rm R}_\odot$.  The remnant of the first hydrostatic core (now a disc with spiral structure), the inner circumstellar disc, and the stellar core are all clearly visible.  The distance scale (in AU), velocity scale (in km~s$^{-1}$), and logarithm of the minimum density (in g~cm$^{-3}$) are given under each panel.  The maximum density is $0.03~{\rm g~cm}^{-3}$ and the logarithm of density is plotted with contours every 0.5 dex. }
\end{figure*}

\begin{figure*}[t]
\epsscale{2.00}
\plotone{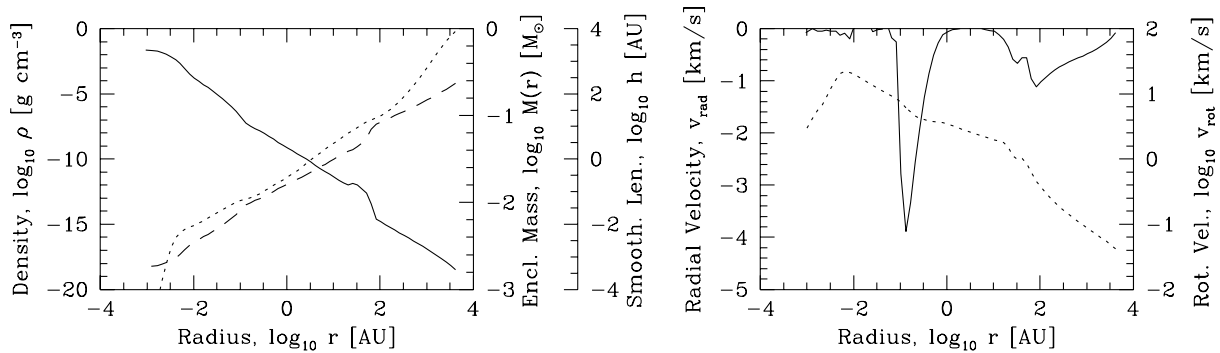}
\caption{The state of the system at the end of the calculation.  The left graph gives the density (solid line) and smoothing length (dashed line) as a function of radius and the mass enclosed within radius $r$ of the center of the stellar core (dotted line).  The right graph gives the radial (solid line) and rotational (dotted line) velocities as functions of radius from the center of the stellar core.  The densities, smoothing lengths and velocities are the mean values in the plane perpendicular to the rotation axis and through the stellar core.  The stellar core ($r<0.004$ AU), inner circumstellar disc ($0.004<r<0.1$ AU), region undergoing the second collapse ($0.1<r<1$ AU), outer circumstellar disc formed from the first core ($1<r<60$ AU), and isothermal collapse region ($r>60$ AU) are clearly visible.}
\end{figure*}

\begin{figure*}[t]
\epsscale{2.00}
\plotone{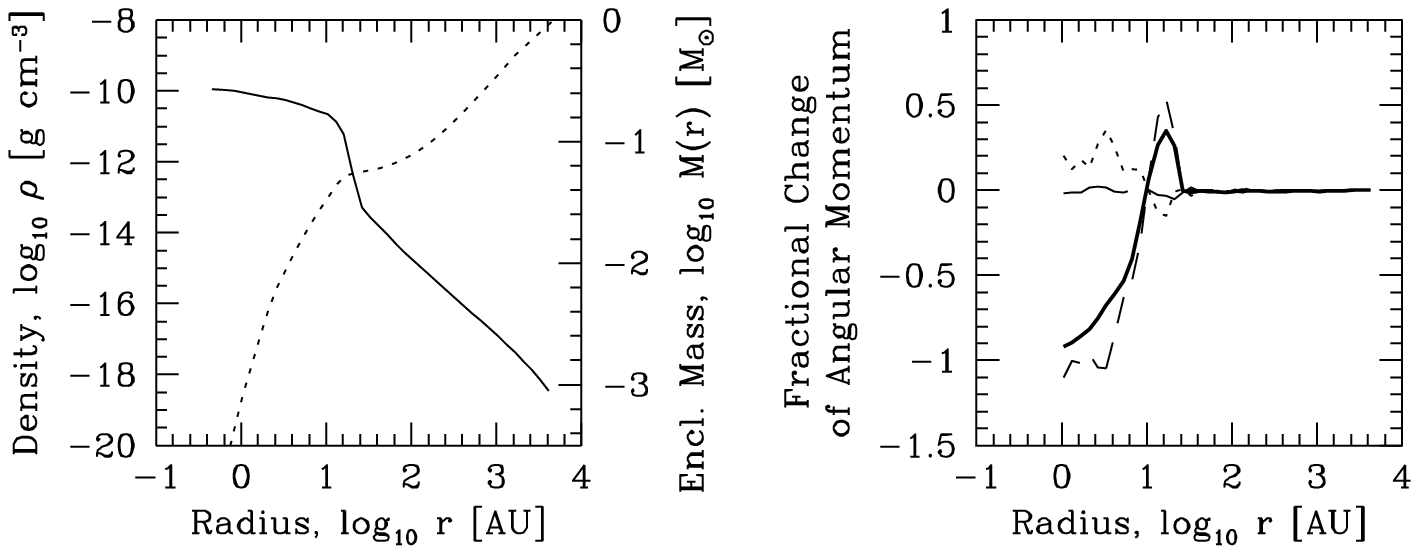}
\caption{The cause of the change in angular momentum during the calculation.  We show the system at $t=1.0234~t_{\rm ff}$.  The first core has undergone approximately 3 rotations since its formation; it has been violently bar-unstable for just 1/4 of a rotation.  The left graph gives the density as a function of radius (solid line) and the mass enclosed within radius $r$ of the center of the first core (dotted line).  The right graph shows the fractional change, since the beginning of the calculation, of the angular momentum of the gas as a function of radius (solid line), and the contributions to this change from gravitational (short-dashed line), pressure (dotted line) and viscous (long-dashed line) torques.  Gravitational torques from the bar instability dominate the loss of angular momentum from the center of the first core; this angular momentum is deposited in the outer parts of the core.  Pressure forces, due to the spiral arms, act to oppose the angular momentum transport.  Viscous transport is negligible.}
\end{figure*}

The evolution of the calculation is as follows (Figure 2).  
The initial collapse is isothermal.  When the density surpasses
$10^{-13}$ g cm$^{-3}$, the gas in the center 
is assumed to become optically thick
to infrared radiation and begins to heat ($t=1.009~t_{\rm ff}$).
The heating stops the collapse at the center and the first hydrostatic core
is formed ($t=1.015~t_{\rm ff}$) with maximum density 
$\approx  2 \times 10^{-11}$ g cm$^{-3}$, mass  
$\approx 0.01 {\rm \ M}_\odot$ ($\approx 3 \times 10^3$ particles), and radius 
$\approx 7$ AU.  As the first core accretes, its maximum
density only slowly increases at first.
However, the first core is rapidly rotating, oblate, and has 
$\beta \approx 0.34 > 0.274$, making it dynamically unstable to the
growth of non-axisymmetric perturbations (e.g.\ Durisen et al.\ 1986).  
At $t \approx 1.023~t_{\rm ff}$, 
after about 3 rotations, the first core becomes violently bar-unstable
and forms trailing spiral arms.  This leads
to a rapid increase in maximum density as angular momentum is removed from
the central regions of the first core (now a disc with spiral structure) 
by gravitational torques 
($t>1.015~t_{\rm ff}$).
When the maximum temperature reaches 2000 K, molecular hydrogen begins to
dissociate, resulting in a rapid second collapse to stellar densities
($t=1.030~t_{\rm ff}$).  The collapse is again halted at a density of 
$\approx 0.007$ g cm$^{-3}$ with the formation of the second hydrostatic, 
or stellar, core.  The initial mass and radius of the stellar core are 
$\approx 0.0015 {\rm \ M}_\odot$ ($\approx 5 \times 10^2$ particles) 
and $\approx 0.8 {\rm \ R}_\odot$, 
respectively.   Finally,
an inner circumstellar disc begins to form around the stellar 
object, within the region undergoing second collapse.  The 
calculation is stopped when the stellar object has a mass of 
$\approx 0.004 {\rm \ M}_\odot$ ($\approx 1.2 \times 10^3$ particles), 
the inner circumstellar disc has extended out to $\approx 0.1$ AU, 
and the outer disc (the remnant of the first hydrostatic core) contains 
$\approx 0.08 {\rm \ M}_\odot$ ($\approx 2.4 \times 10^4$ particles) 
and extends out to $\approx 60$ AU.  Note that the massive outer 
disc forms {\it before} the stellar core.
This final state is illustrated in Figures 3 and 4.
If the calculation was evolved further, the inner circumstellar disc would 
continue to grow in radius as gas with higher angular momentum fell in.
Eventually, the inner disc would meet the outer disc with only a
small molecular dissociation region between the two.

The value of $\beta$ during the second collapse, or after the stellar 
core and disc have formed, never exceeds the value necessary for 
dynamic instability to non-axisymmetric perturbations ($\beta=0.274$).
Therefore, unlike in the calculations of Bonnell \& Bate (1994), fragmentation 
to form a close binary system cannot occur.  A similar result is obtained
with $\Omega=3.4 \times 10^{-14}$ rad s$^{-1}$ and $\beta=0.001$ initially.

Since whether or not fragmentation can occur depends critically on the 
value of $\beta$, it is important that there has not been 
significant spurious transport of angular momentum by artificial viscosity 
or the diffusion of SPH particles.  As SPH is a Lagrangian technique, 
it is relatively simple to keep track of the gravitational, pressure, 
and viscous torques on each particle.  Thus, we can determine
the cause of the change in angular momentum of a volume of gas 
between any two times.  As one might expect, we find that there is 
virtually no angular momentum transport during dynamic
collapse.  Viscous evolution is more likely during the quasistatic
evolution of the first core between its formation and the second collapse.
In Figure 5, we consider the cause of the angular momentum transport during
the quasistatic first core phase.  Viscous evolution, which has presumably 
been active during the $\approx 3$ rotations that the first core made before 
it became violently bar-unstable, is hardly detectable.  By contrast, it takes
only 1/4 of a rotation for gravitational torques to remove most of the 
angular momentum from the gas in the center of the core.

The absence of fragmentation could also be due to the value 
of $\gamma$ during the second collapse being overestimated.  This would lead
to the value of beta that is achieved at the end of the collapse being
underestimated.
To determine the dependence of the results on $\gamma$, we performed
calculations with $\gamma$ as low as 1.05.  This corresponds to an 
{\it isothermal}
collapse but with the mean molecular weight still {\it decreasing} due to the 
dissociation of molecular hydrogen.  In this extreme case $\beta$ does 
briefly exceed 0.274 as the stellar core begins to form, but drops below 
the critical value again over approximately one rotation as the core 
increases in mass and before non-axisymmetric perturbations are able to
grow significantly.  The inner disc that forms around the stellar core also
has a brief period with $\beta > 0.274$ until more mass is accreted.  
Weak, transient spiral arms are visible in the disc over several rotations, 
however, there is no tendency for fragmentation.

\section{Conclusions}

Larson (1969) performed the first hydrodynamical calculations of the
collapse of a molecular cloud core to stellar densities using a 
spherically-symmetric, one-dimensional code.  After nearly 30 years, 
it is now possible to perform three-dimensional calculations.
Using SPH, we have followed the 
three-dimensional collapse of a molecular cloud core over 17 orders of 
magnitude in density contrast and 7 orders of magnitude in spatial 
extent, resolving both the isothermal and second collapse phases, 
and the formation of the first and second hydrostatic cores.

Using this technique, we have begun a study of whether or not close binary
stellar systems may be formed during the second collapse phase
of a molecular cloud core, as proposed by
Bonnell \& Bate (1994).  We find that, if the first hydrostatic core is
formed with sufficient rotational energy to make it dynamically unstable
to the growth of non-axisymmetric perturbations, fragmentation during the
second collapse is prevented because of the removal of angular momentum from
the central regions of the first core via gravitational torques.  
The formation of close binaries during the second collapse phase {\it may 
be possible} for molecular cloud cores with different initial 
conditions, {\it especially} if they are rotating slowly enough that the first
core is rotationally stable.  Different initial conditions will be 
investigated in a subsequent paper.



\begin{thebibliography}{}
\bibitem[Bate, Bonnell \& Price 1995]{BBP95} Bate, M. R., Bonnell, I. A., \& Price, N. M. 1995, \mnras, 277, 362
\bibitem[Bate \& Burkert 1997]{BatBur97} Bate, M.R., \& Burkert, A. 1997, \mnras, 288, 1060
\bibitem[Benz 1990]{Benz90} Benz, W. 1990, in The Numerical Modeling of Nonlinear Stellar Pulsations: Problems and Prospects, ed. J. R. Buchler, (Dordrecht: Kluwer), 269
\bibitem[Benz et al.\ 1990]{Benzetal1990} Benz, W., Bowers, R. L., Cameron, A. G. W., \& Press, W. 1990, \apj, 348, 647
\bibitem[Bonnell \& Bate 1994]{BonBat94} Bonnell, I. A., \& Bate, M. R. 1994, \mnras, 271, 999
\bibitem[Boss 1986]{Boss86} Boss, A. P. 1986, \apjs, 62, 519
\bibitem[Boss 1988]{Boss88} Boss, A. P. 1988, \apj, 331, 370
\bibitem[Boss 1989]{Boss89} Boss, A. P. 1989, \apj, 346, 336
\bibitem[Burkert \& Bodenheimer 1993]{BurBod93} Burkert, A, \& Bodenheimer, P. 1993, \mnras, 264, 798
\bibitem[Durisen et al.\ 1986]{DGTB86} Durisen, R. H., Gingold, R. A., Tohline, J. E., Boss, A. P., 1986, \apj, 305, 281
\bibitem[Larson 1969]{Larson69} Larson, R. B. 1969, \mnras, 145, 271
\bibitem[Monaghan \& Gingold 1983]{MonGin83} Monaghan, J. J., \& Gingold, R. A. 1983, J. Comput. Phys., 52, 374
\bibitem[Tohline 1982]{Tohline82} Tohline, J. E. 1982, Fund. Cos. Phys., 8, 1
\bibitem[Truelove et al.\ 1998]{Trueloveetal1998} Truelove, J. K., Klein, R. I., McKee, C. F., Holliman, J. H., II, Howell, L. H., Greenough, J. A., \& Woods, D. T. 1998, \apj, 495, 821
\bibitem[Whitworth 1998]{Whitworth98} Whitworth, A. P. 1998, \mnras, 296, 442
\bibitem[Winkler \& Newman 1980]{WinNew80a} Winkler, K.-H. A., Newman, M. J., 1980, \apj, 236, 201
\bibitem[Winkler \& Newman 1980]{WinNew80b} Winkler, K.-H. A., Newman, M. J., 1980, \apj, 238, 311

\end{thebibliography}
\end{document}